\documentclass[12pt]{article}
\begin{document}

\title{\bf Gravitational Perfect Fluid Collapse With Cosmological Constant}

\author{M. Sharif \thanks{msharif@math.pu.edu.pk} and Zahid Ahmad
\thanks{zahid$\_$rp@yahoo.com}\\
Department of Mathematics, University of the Punjab,\\
Quaid-e-Azam Campus, Lahore-54590, Pakistan.}

\date{}
\maketitle

\begin{abstract}
In this paper, the effect of a positive cosmological constant on
spherically symmetric collapse with perfect fluid has been
investigated. The matching conditions between static exterior and
non-static interior spacetimes are given in the presence of a
cosmological constant. We also study the apparent horizons and their
physical significance. It is concluded that the cosmological
constant slows down the collapse of matter and hence limit the size
of the black hole. This analysis gives the generalization of the
dust case to the perfect fluid. We recover the results of the dust
case for $p=0$.
\end{abstract}


\section{Introduction}

The cosmological constant is an energy associated with the vacuum,
i.e., with empty space. The inclusion of the non-zero cosmological
constant into the Einstein field equations has been discussed
several times in the past for theoretical and observational reasons
[1]. First, it has been introduced by Einstein to save the universe
from expanding and rejected by him after expansion has been
discovered by Hubble. The results of type Ia supernova [2,3] show
that the universe is accelerating rather than decelerating. These
results suggest that our universe can have a non-zero cosmological
constant. Another analysis [4] of the peculiar motion of low-red
shift galaxies give further evidence for the possibility of finite
cosmological constant. These results have increased interest to
study the properties of the universe with a non-zero cosmological
constant. The physical applications of a cosmological constant are
huge, restricting not only the growth of the universe but also the
structure formation and age problems. The cosmological constant
affects the properties of spacetime and matter. Since the metric of
the spacetime and the stress-energy tensor of matter are related
through the Einstein field equations, the effects of a cosmological
constant can be analyzed by specifying the metric and the
stress-energy tensor. We study gravitational collapse to see such
effects.

Gravitational collapse is one of the important issues in General
Relativity. This theory predicts solutions with singularities and
such solutions can be produced by the gravitational collapse of
non-singular, asymptotically flat initial data [5-7]. Spacetime
singularities can be classified into two kinds whether they can be
observed or not. A spacetime singularity is said to be naked when it
is observable to local or distant observer. If such singularity can
reach the neighboring or asymptotic regions of spacetime, the
singularity is called locally or globally naked singularity. A
spacetime singularity which can not be observed is called a black
hole. Is such a singularity formed in our universe? Penrose [8]
proposed so-called the cosmic censorship conjecture to resolve this
problem. According to this conjecture, the singularities that appear
in the gravitational collapse are always covered by an event
horizon. This conjecture has provided a strong motivation for
researchers in this field. The compact stellar objects such as white
dwarf and neutron star are formed by a period of gravitational
collapse. It is interesting to consider the appropriate geometry of
interior and exterior regions and determine proper junction
conditions which allow the matching of these regions.

Most of the problems related to gravitational collapse have been
discussed by considering spherically symmetric system. The
gravitational collapse of dust was first shown by Oppenheimer and
Snyder [9]. They studied collapse by considering static
Schwarzschild in the exterior and Friedman like solution in the
interior. Many people [10-14] extended the above study of collapse
by taking an appropriate geometry of interior and exterior regions.
Markovic and Shapiro [15] generalized the work done by Oppenheimer
and Snyder [9] in the presence of positive cosmological constant.
Later, Lake [16] generalized the results of Markovic and Shapiro
[15] for both positive and negative cosmological constant. Cissoko
et al. [17] discussed explicitly gravitational dust collapse with
positive cosmological constant. Recently, the same work has been
generalized by Ghosh and Deshkar [18] for higher dimensional dust
collapse with cosmological constant.

In this paper, we discuss the gravitational collapse with
cosmological constant for perfect fluid case. It is verified that
our results reduce to the dust case as given by Cissoko et al. [17].
The paper is outlined as follows. In section 2, we give the junction
conditions between a static and a non-static spherically symmetric
spacetimes. Section 3 yields the spherically symmetric perfect fluid
solution of the Einstein field equations with a cosmological
constant. In section 4, we discuss the solution with some
assumptions. Section 5 is devoted to investigate the apparent
horizons and the role of the cosmological constant. Finally, we
summarize the results in section 6.

\section{Junction Conditions}

We consider a timelike $3D$ hypersurface $\Sigma$, which divides
$4D$ spacetime into two regions interior and exterior spacetimes,
denoted by $V^+$ and $V^-$ respectively. For the interior spacetime,
we consider spherically symmetric system given by
\begin{equation}
ds_{-}^2=dt^2-X^2dr^2-Y^2(d\theta^2+\sin^2{\theta}d\phi^2),
\end{equation}
where $X$ and $Y$ are functions of $t$ and $r$ only. For the
exterior spacetime, we take the Schwarzschild-de Sitter metric,
\begin{equation}
ds_{+}^2=FdT^2-\frac{1}{F}dR^2-R^2(d\theta^2+\sin^2{\theta}d\phi^2),
\end{equation}
where
\begin{equation}
F(R)=1-\frac{2M}{R}-\frac{\Lambda}{3}R^2,
\end{equation}
$M$ is a constant and $\Lambda$ is the cosmological constant.
According to the junction conditions [19,20], it is assumed that the
first and second fundamental forms from the interior and the
exterior spacetimes are the same. These conditions can be expressed
as\\
\par\noindent
(i) The continuity of the first fundamental form over $\Sigma$ gives
\begin{equation}
(ds_{-}^2)_{\Sigma}=(ds_{+}^2)_{\Sigma}=ds_{\Sigma}^2.
\end{equation}
(ii) The continuity of the second fundamental form over $\Sigma$
gives
\begin{equation}
[K_{ab}]=K_{ab}^+-K_{ab}^-=0,\quad (a,b=0,2,3),
\end{equation}
where $K_{ab}$, the extrinsic curvature, is given by
\begin{equation}
K_{ab}^\pm=-n_{\sigma}^\pm(\frac{\partial^2x_\pm^\sigma}{\partial\xi^a\partial\xi^b}
+\Gamma^\sigma_{\mu\nu}\frac{\partial x_\pm^\mu}{\partial\xi^a}
\frac{\partial x_\pm^\nu}{\partial\xi^b}),\quad (\sigma, \mu, \nu=0,1,2,3).
\end{equation}
Here the Christoffel symbols $\Gamma^\sigma_{\mu\nu}$ are calculated
from the interior or exterior metrics (1) or (2), $n_{\mu}^\pm$ are
the components of outward unit normals to $\Sigma$ in the
coordinates $x_\pm^\sigma$. The equations of hypersurface $\Sigma$
in the coordinates $x_\pm^\sigma$ are written as
\begin{eqnarray}
f_{-}(r,t)&=&r-r_{\Sigma}=0,\\
f_{+}(R,T)&=&R-R_{\Sigma}(T)=0,
\end{eqnarray}
where $r_{\Sigma}$ is a constant.

Using Eq.(7) in (1), the metric on $\Sigma$ takes the form
\begin{equation}
(ds_{-}^2)_{\Sigma}=dt^2-[Y(r_{\Sigma},t)]^2(d\theta^2+\sin^2{\theta}d\phi^2).
\end{equation}
Similarly, Eqs.(2) and (8) yield
\begin{equation}
(ds_{+}^2)_{\Sigma}=[F(R_{\Sigma})-\frac{1}{F(R_{\Sigma})}
(\frac{dR_{\Sigma}}{dT})^2]dT^2-{R_{\Sigma}}^2(d\theta^2
+\sin^2{\theta}d\phi^2),
\end{equation}
where we assume that
\begin{equation}
F(R_{\Sigma})-\frac{1}{F(R_{\Sigma})} (\frac{dR_{\Sigma}}{dT})^2>0
\end{equation}
so that $T$ is a timelike coordinate. From Eqs.(4), (9) and (10), it
follows that
\begin{equation}
R_{\Sigma}=Y(r_\Sigma,t),
\end{equation}
\begin{equation}
[F(R_{\Sigma})-\frac{1}{F(R_{\Sigma})}
(\frac{dR_{\Sigma}}{dT})^2]^\frac{1}{2} dT=dt.
\end{equation}
Now from Eqs.(7) and (8), the outward unit normals in $V^{-}$ and
$V^{+}$, respectively, are given by
\begin{eqnarray}
n_\mu^-&=&(0,X(r_\Sigma,t),0,0),\\
n_\mu^+&=&(-\dot{R_\Sigma},\dot{T},0,0),
\end{eqnarray}
where dot means differentiation with respect to $t$. The components
of the extrinsic curvature $K_{ab}^\pm$ are
\begin{eqnarray}
K_{00}^-&=&0,\\
K_{22}^-&=&\csc^2{\theta}K_{33}^-=(\frac{YY'}{X})_\Sigma,\\
K_{00}^+&=&(\dot{R}\ddot{T}-\dot{T}\ddot{R}-\frac{F}{2}
\frac{dF}{dR}{\dot{T}}^3+\frac{3}{2F}\frac{dF}{dR}\dot{T}{\dot{R}}^2)_\Sigma ,\\
K_{22}^+&=&\csc^2{\theta}K_{33}^+=(FR\dot{T})_\Sigma.
\end{eqnarray}
The continuity of the extrinsic curvature gives
\begin{eqnarray}
K_{00}^+&=&0,\\
K_{22}^-&=&K_{22}^+.
\end{eqnarray}
When we use Eqs.(16)-(21) along with Eq.(3), the junction
conditions turn out to be
\begin{eqnarray}
(X\dot{Y}'-\dot{X}Y')_\Sigma=0,\\
M=(\frac{Y}{2}-\frac{\Lambda}{6}Y^3+\frac{Y}{2}{\dot{Y}}^2
-\frac{Y}{2X^2}{Y'}^2)_\Sigma.
\end{eqnarray}

\section{Solution of the Field Equations}

The Einstein field equations for perfect fluid with cosmological
constant are given by
\begin{equation}
R_{\mu\nu}=8\pi[(\rho+p)u_{\mu}u_{\nu}
+\frac{1}{2}(p-\rho)g_{\mu\nu}]-\Lambda g_{\mu\nu},
\end{equation}
where $\rho$ is the energy density, $p$ is the pressure and
$u_{\mu}=\delta_\mu^0$ is the four-velocity in co-moving
coordinates. These equations for the line element (1) take the form
\begin{eqnarray}
R_{00}&=&-\frac{\ddot{X}}{X}-2\frac{\ddot{Y}}{Y}=4\pi(\rho+3p)-\Lambda,\\
R_{11}&=&-\frac{\ddot{X}}{X}-2\frac{\dot{X}}{X}\frac{\dot{Y}}{Y}
+\frac{2}{X^2}(\frac{Y''}{Y}-\frac{X'}{X}\frac{Y'}{Y})=4\pi(p-\rho)-\Lambda,\\
R_{22}&=&-\frac{\ddot{Y}}{Y}-(\frac{\dot{Y}}{Y})^2-\frac{\dot{X}}{X}\frac{\dot{Y}}{Y}
+\frac{1}{X^2}[\frac{Y''}{Y}+(\frac{Y'}{Y})^2-\frac{X'}{X}\frac{Y'}{Y}
-(\frac{X}{Y})^2]\nonumber\\
&=&4\pi(p-\rho)-\Lambda,\\
R_{33}&=&\sin^2{\theta}R_{22},\\
R_{01}&=&-2\frac{\dot{Y}'}{Y}+2\frac{\dot{X}}{X}\frac{Y'}{Y}=0.
\end{eqnarray}
Now we solve these equations. When we integrate Eq.(29) w.r.t. $t$,
we get
\begin{equation}
X=\frac{Y'}{W},
\end{equation}
where $W=W(r)$ is an arbitrary function of $r$ . Using this value of
$X$ in Eqs.(25)-(27), it follows that
\begin{equation}
2\frac{\ddot{Y}}{Y}+(\frac{\dot{Y}}{Y})^2+\frac{1-W^2}{Y^2}=\Lambda-8\pi
p.
\end{equation}
Integrating this equation w.r.t. $t$, it turns out that
\begin{equation}
\dot{Y}^2=W^2-1+2\frac{m}{Y}+(\Lambda-8\pi p)\frac{Y^2}{3},
\end{equation}
where $m=m(r)$ is an arbitrary function of $r$ and is related to the
mass of the collapsing system. When we use Eqs.(30) and (32) in
(25), we obtain
\begin{equation}
m'=4\pi(\rho+p)Y^2Y'-\frac{1}{3}8\pi p'Y^3.
\end{equation}
For physical reasons, we assume that density and pressure are
non-negative. Integration of Eq.(33) w.r.t $r$ yields
\begin{equation}
m(r)=4\pi\int^r_0(\rho+p)Y^2Y'dr-\frac{8\pi}{3}\int^r_0p'Y^3dr.
\end{equation}
Here we take constant of integration to be zero. The function
$m(r)$ must be positive, because $m(r)<0$ implies negative mass
which is not physical. Using Eqs.(30) and (32) into the junction
condition (23), we obtain
\begin{equation}
M=m-\frac{4\pi p}{3}Y^3.
\end{equation}
We see from Eq.(3) that the exterior spacetime becomes the
Schwarzschild spacetime for $\Lambda=0$ and $M$ as the total energy
inside the surface $\Sigma$ due to its Newtonian asymptotic
behaviour. The total energy $\tilde{M}(r,t)$ up to a radius $r$ at
time $t$ inside the hypersurface $\Sigma$ can be evaluated by using
the definition of the mass function [10,17]. For the metric (1), it
takes the following form
\begin{equation}
\tilde{M}(r,t)=\frac{1}{2}Y^3{R^{23}}_{23}=\frac{1}{2}Y[1-(\frac{Y'}{X})^2+\dot{Y}^2].
\end{equation}
Using Eqs.(30) and (32) in Eq.(36), it follows that
\begin{equation}
\tilde{M}(r,t)=m(r)+(\Lambda-8\pi p)\frac{Y^3}{6}.
\end{equation}

\section{Solution With $W(r)=1$}

In this section, we consider the case $\Lambda-8\pi p>0$ and the
condition
\begin{equation}
W(r)=1.
\end{equation}
Using Eqs.(30), (32) and (38), we obtain the analytic solutions in
closed form as
\begin{eqnarray}
Y(r,t)&=&(\frac{6m}{\Lambda-8\pi
p})^{\frac{1}{3}}\sinh^{\frac{2}{3}}\alpha(r,t),\\
X(r,t)&=&(\frac{6m}{\Lambda-8\pi
p})^{\frac{1}{3}}[\{\frac{m'}{3m}+\frac{16\pi m p'}{(\Lambda-8\pi
p)^2}\}\sinh\alpha(r,t)\nonumber\\
&+&\{\frac{-8\pi p'(t_0-t)}{\sqrt{3(\Lambda-8\pi
p)}}+{t_0}'\sqrt{\frac{\Lambda-8\pi
p}{3}}\}\cosh\alpha(r,t)]\sinh^{\frac{-1}{3}}\alpha(r,t),\nonumber\\
\end{eqnarray}
where
\begin{equation}
\alpha(r,t)=\frac{\sqrt{3(\Lambda-8\pi p)}}{2}[t_0(r)-t].
\end{equation}
Here $t_0(r)$ is an arbitrary function of $r$. In the limit
$p\rightarrow\frac{\Lambda}{8\pi}$, the above solution corresponds
to the Tolman-Bondi solution [21]
\begin{eqnarray}
\lim_{ p\rightarrow\frac{\Lambda}{8\pi}}Y(r,t)
&=&[\frac{9m}{2}(t_0-t)^2]^{\frac{1}{3}},\\
\lim_{p\rightarrow\frac{\Lambda}{8\pi}}X(r,t)
&=&\frac{m'(t_0-t)+2mt_0'}{[6m^2(t_0-t)]^\frac{1}{3}}.
\end{eqnarray}

\section{Apparent Horizons}

When the boundary of trapped two spheres is formed, we obtain the
apparent horizon. Here we find this boundary of the trapped two
spheres whose outward normals are null. For Eq.(1), this is given as
follows
\begin{equation}
g^{\mu\nu}Y_{,\mu} Y_{,\nu}=\dot{Y}^2-(\frac{Y'}{X})^2=0.
\end{equation}
Using Eqs.(30) and (32) in Eq.(44), we obtain
\begin{equation}
(\Lambda-8\pi p)Y^3-3Y+6m=0.
\end{equation}
The solutions of the above equation for $Y$ give the apparent
horizons. For $\Lambda=8\pi p$, it becomes the Schwarzschild
horizon, i.e., $Y=2m$. When $m=0,~ p=0$, it yields the de-Sitter
horizon $Y=\sqrt{\frac{3}{\Lambda}}$. The case
$3m<\frac{1}{\sqrt{\Lambda-8\pi p}}$ leads to two horizons
\begin{eqnarray}
Y_1&=&\frac{2}{\sqrt{\Lambda-8\pi p}}\cos\frac{\psi}{3},\\
Y_2&=&-\frac{2}{\sqrt{\Lambda-8\pi
p}}(\cos\frac{\psi}{3}-\sqrt{3}\sin\frac{\psi}{3}),
\end{eqnarray}
where
\begin{equation}
\cos\psi=-3m\sqrt{\Lambda-8\pi p}.
\end{equation}
For $m=0$, it follows from Eqs.(46) and (47) that
$Y_1=\sqrt{\frac{3}{\Lambda-8\pi p}}$ and $Y_2=0$. $Y_1$ is called
the cosmological horizon and $Y_2$ is referred to be black hole
horizon which can be generalized for $m\neq0$ and $\Lambda\neq8\pi
p$ respectively [22]. It is mentioned here that both horizons
coincide for $3m=\frac{1}{\sqrt{\Lambda-8\pi p}}$, i.e.,
\begin{equation}
Y_1=Y_2=\frac{1}{\sqrt{\Lambda-8\pi p}}=Y
\end{equation}
which gives a single horizon. It is obvious that the range for the
cosmological horizon and the black hole horizon turns out to be
\begin{equation}
0\leq Y_2\leq \frac{1}{\sqrt{\Lambda-8\pi p}}\leq Y_1\leq
\sqrt{\frac{3}{\Lambda-8\pi p}}.
\end{equation}
The black hole horizon has its largest proper area $4\pi
Y^2=\frac{4\pi}{\Lambda}$ and the cosmological horizon has an area
between $\frac{4\pi}{\Lambda-8\pi p}$ to $\frac{12\pi}{\Lambda-8\pi
p}$. For $3m>\frac{1}{\sqrt{\Lambda-8\pi p}}$, there are no
horizons. The formation time of the apparent horizon can be
calculated with the help of Eqs.(38), (39) and (45) and is given by
\begin{equation}
t_n=t_0-\frac{2}{\sqrt{3(\Lambda-8\pi
p)}}\sinh^{-1}(\frac{Y_n}{2m}-1)^\frac{1}{2},\quad (n=1,2).
\end{equation}
In the limit $p\rightarrow\frac{\Lambda}{8\pi}$, we obtain the
result corresponding to Tolman-Bondi [21]
\begin{equation}
t_{ah}=t_0-\frac{4}{3}m.
\end{equation}
From Eq.(51), it can be seen that both the black hole horizon and
the cosmological horizon form earlier than the singularity
$t=t_0$. From Eq.(51), it follows that
\begin{equation}
\frac{Y_n}{2m}=\cosh^2\alpha_n.
\end{equation}
Eq.(50) yields that $Y_1\geq Y_2$ and also Eq.(51) gives $t_1\leq
t_2$, i.e., cosmological horizon forms earlier than the formation
of the black hole horizon. To see the time difference between the
formation of the cosmological horizon and singularity and the
formation of the black hole horizon and singularity respectively,
using Eqs.(46)-(48), we need to calculate the following quantities
\begin{eqnarray}
\frac{d(\frac{Y_1}{2m})}{dm}&=&\frac{1}{m}(-\frac{\sin\frac{\psi}{3}}{\sin\psi}
+\frac{3\cos\frac{\psi}{3}}{\cos\psi})<0,\\
\frac{d(\frac{Y_2}{2m})}{dm}&=&\frac{1}{m}(-\frac{\sin\frac{(\psi+4\pi)}{3}}{\sin\psi}
+\frac{3\cos\frac{(\psi+4\pi)}{3}}{\cos\psi})>0.
\end{eqnarray}
We define the time difference between the formation of singularity
and the apparent horizon, denoted by $\tau$ as follows
\begin{equation}
\tau_n=t_0-t_n.
\end{equation}
It follows from Eq.(53) that
\begin{equation}
\frac{d\tau_n}{d(\frac{Y_n}{2m})}=\frac{1}{\sinh\alpha_n\cosh\alpha_n\sqrt{3(\Lambda
-8\pi p)}}.
\end{equation}
Using Eqs.(54) and (57), it turns out that
\begin{equation}
\frac{d\tau_1}{dm}=\frac{d\tau_1}{d(\frac{Y_1}{2m})}\frac{d(\frac{Y_1}{2m})}{dm}
=\frac{1}{m\sqrt{3(\Lambda-8\pi
p)}\sinh\alpha_1\cosh\alpha_1}(-\frac{\sin\frac{\psi}{3}}{\sin\psi}
+\frac{3\cos\frac{\psi}{3}}{\cos\psi})<0.
\end{equation}
This shows that $\tau_1$ decreases and hence the time difference
between the formation of singularity and cosmological horizon
decreases. From Eqs.(55) and (57), it follows that
\begin{equation}
\frac{d\tau_2}{dm} =\frac{1}{m\sqrt{3(\Lambda-8\pi
p)}\sinh\alpha_2\cosh\alpha_2}(-\frac{\sin\frac{(\psi+4\pi)}{3}}{\sin\psi}
+\frac{3\cos\frac{(\psi+4\pi)}{3}}{\cos\psi})>0.
\end{equation}
This implies that $\tau_2$ increases which means that the time
difference between the formation of singularity and black hole
horizon increases.

\section{Conclusion}

In this paper, we have studied the gravitational collapse of a
perfect fluid in the presence of a cosmological constant. The
effects of the cosmological constant on gravitational collapse have
been discussed in the following two ways.

Firstly, the cosmological constant plays the role of repulsive
force, i.e., it slows down the collapsing process. The cosmological
term behaves like a Newtonian potential given by
$\phi=\frac{1}{2}(1-g_{00})$. Using Eqs.(12) and (35) for the
exterior metric, the Newtonian potential takes the following form
\begin{equation}
\phi(R)=\frac{m}{R}+(\Lambda-8\pi p)\frac{R^2}{6}.
\end{equation}
The corresponding Newtonian force turns out to be
\begin{equation}
F=-\frac{m}{R^2}+(\Lambda-8\pi p)\frac{R}{3}
\end{equation}
which vanishes for $R=\frac{1}{\sqrt{\Lambda-8\pi p}}$ and
$m=\frac{1}{3\sqrt{\Lambda-8\pi p}}$. Thus the force becomes
repulsive/attarctive for larger/smaller mass and radius respectively
than these values. This means that the size of the black hole can be
visualized by comparing the repulsive and attractive forces. The
repulsive force generates from the cosmological constant for
$\Lambda>8\pi p$. It is worth mentioning here that for the perfect
fluid $\Lambda$ can play the role of a repulsive force only for
$\Lambda>8\pi p$ while in the dust case this is true for all values
of $\Lambda>0$. From Eq.(32), the rate of collapse turns out to be
\begin{equation}
\ddot{Y}= -\frac{m}{Y^2}+(\Lambda-8\pi p)\frac{Y}{3}.
\end{equation}
For collapsing process, the force should be attractive, i.e., the
acceleration should be negative which implies that
$Y<(\frac{3m}{\Lambda-8\pi p})^\frac{1}{3}$. Thus Eq.(62) shows that
the cosmological constant slows down the collapsing process if
$\Lambda>8\pi p$. This means that, for $p>\frac{\Lambda}{8\pi}$, the
force becomes attractive and hence the cosmological constant does
not slow down the collapsing process.

Secondly, there are two physical horizons instead of one due to the
presence of the term $\Lambda-8\pi p$, i.e., the black hole horizon
and the cosmological horizon respectively. The cosmological constant
influences the time difference between the formation of the apparent
horizon and singularity. We find that the cosmological constant
affects the process of collapse and hence it limits the size of the
black hole. In perfect fluid case, these results are valid only for
$\Lambda>8\pi p$ while in the dust case these are valid for all
$\Lambda>0$. Thus we conclude that the pressure term creates a bound
for the cosmological constant to act as a repulsive force. It is
mentioned here that if we take $p=0$, the results reduce to the dust
case [17].

\newpage

\vspace{0.5cm}

\vspace{0.5cm}

{\bf Acknowledgment}

\vspace{0.5cm}

This work has been completed with the financial support of the
Higher Education Commission Islamabad, Pakistan through the {\it
Indigenous PhD 5000 Fellowship Program Batch-I}. We are also
thankful for the anonymous referee for his useful comments.

\vspace{0.5cm}

{\bf \Large References}

\begin{description}

\item{[1]} Cohn, J.D.: Astrophysics. J. Suppl. {\bf 259}(1998)213.

\item{[2]} Perlmutter, S., et al.: Nature {\bf 391}(1998)51.

\item{[3]} Reiss, A.G, et al.: Astron. J. {\bf 116}(1998)1009.

\item{[4]} Zehavi, I. and Dekel, A.: Nature {\bf 401}(1999)252.

\item{[5]} Penrose, R.: Phys. Rev. Lett. {\bf 14}(1965)57.

\item{[6]} Hawking, S.W.: {\it Proc. R. Soc. London} {\bf A300}(1967)187.

\item{[7]} Hawking, S.W. and Penrose, R.: {\it Proc. R. Soc. London} {\bf A314}(1970)529.

\item{[8]} Penrose, R.: Riv. Nuovo Cimento {\bf 1}(1969)252.

\item{[9]} Oppenheimer, J.R. and Snyder, H.: Phys. Rev. {\bf 56}(1939)455.

\item{[10]} Misner, C.W. and Sharp, D.: Phys. Rev. {\bf D136}(1964)b571.

\item{[11]} Ghosh, S.G. and Deshkar, D.W.: Int. J. Mod. Phys. {\bf D12}(2003)317.

\item{[12]} Ghosh, S.G. and Deshkar, D.W.: Gravitation and Cosmology {\bf 6}(2000)1.

\item{[13]} Debnath, U., Nath, S. and Chakraborty, S.: Mon. Not. R. Astron.
Soc. {\bf 369}(2006)1961.

\item{[14]} Debnath, U., Nath, S. and Chakraborty, S.: Gen. Relativ. Grav. {\bf 37}(2005)215.

\item{[15]} Markovic, D. and Shapiro, S.L.: Phys. Rev. {\bf D61}(2000)084029.

\item{[16]} Lake, K.: Phys. Rev. {\bf D62}(2000)027301.

\item{[17]} Cissoko, M., Fabris, J.C., Gariel, J., Denmat, G.L. and Santos, N.O.:
 arXiv:gr-qc/9809057.

\item{[18]} Ghosh, S.G. and Deshkar, D.W.: arXiv:gr-qc/0607142

\item{[19]} Santos, N.O.: Phys. Lett. {\bf A106}(1984)296.

\item{[20]} Israel, W.: Nuovo Cimento {\bf 44B}(1966)1.

\item{[21]} Eardley, D.M. and Smarr, L.: Phys. Rev. {\bf D19}(1979)2239.

\item{[22]} Hayward, S.A., Shiromizu, T. and Nakao, K.: Phys. Rev. {\bf D49}(1994)5080.

\end{description}
\end{document}